\documentclass[prb,twocolumn,showpacs,preprintnumbers,amsmath,amssymb,superscriptaddress]{revtex4}
\usepackage{graphicx}% Include figure files
\usepackage{dcolumn}% Align table columns on decimal point
\usepackage{bm}% bold math

\begin{document}

\title{Quasi-classical cyclotron resonance of Dirac fermions in highly doped graphene}

\author{A. M. Witowski}
\affiliation{Institute of Experimental Physics, University of Warsaw, Ho\.{z}a
69,  00-681 Warsaw, Poland}
\author{M. Orlita}\email{milan.orlita@lncmi.cnrs.fr}
\affiliation{Laboratoire National des Champs Magn\'etiques Intenses,
CNRS-UJF-UPS-INSA, 25, avenue des Martyrs, 38042 Grenoble, France}
\affiliation{Institute of Physics, Charles University, Ke Karlovu 5,
121~16 Praha 2, Czech Republic}
\affiliation{Institute of Physics, v.v.i., ASCR,
Cukrovarnick\'{a} 10, CZ-162 53 Praha 6, Czech Republic}
\author{R. St\c{e}pniewski}
\affiliation{Institute of Experimental Physics, University of Warsaw, Ho\.{z}a
69,  00-681 Warsaw, Poland}
\author{A. Wysmo\l{}ek}
\affiliation{Institute of Experimental Physics, University of Warsaw, Ho\.{z}a
69,  00-681 Warsaw, Poland}
\author{J.~M.~Baranowski}
\affiliation{Institute of Experimental Physics, University of Warsaw, Ho\.{z}a
69,  00-681 Warsaw, Poland}
\affiliation{Institute of Electronic Materials Technology, Wolczynska 133, 01-919 Warsaw, Poland}
\author{W. Strupi\'{n}ski}
\affiliation{Institute of Electronic Materials Technology, Wolczynska 133, 01-919 Warsaw, Poland}
\author{C. Faugeras}
\affiliation{Laboratoire National des Champs Magn\'etiques Intenses,
CNRS-UJF-UPS-INSA, 25, avenue des Martyrs, 38042 Grenoble, France}
\author{G. Martinez}
\affiliation{Laboratoire National des Champs Magn\'etiques Intenses,
CNRS-UJF-UPS-INSA, 25, avenue des Martyrs, 38042 Grenoble, France}
\author{M. Potemski}
\affiliation{Laboratoire National des Champs Magn\'etiques Intenses,
CNRS-UJF-UPS-INSA, 25, avenue des Martyrs, 38042 Grenoble, France}
\date{\today}

\begin{abstract}
Cyclotron resonance in highly doped graphene has been explored using infrared magneto-transmission. Contrary to previous work, which only focused on the magneto-optical properties of graphene in the quantum regime, here we study the quasi-classical response of this system. We show that it has a character of classical cyclotron resonance, with an energy which is linear in the applied magnetic field and with an effective cyclotron mass defined by the position of the Fermi level $m=E_F/v_F^2$.
\end{abstract}


\maketitle

\section{Introduction}

The cyclotron motion of charge carriers with a well defined effective mass $m$ and the related resonant absorption of light at the cyclotron frequency
$\omega_c=|e|B/m$ are basic physical phenomena representative of the magneto-optical response of conventional condensed-matter systems. Well-established and often applied, cyclotron resonance (CR) became through the years a common technique routinely used in solid-state physics, inherently connected with the effective 
mass of elementary electronic excitations.\cite{CohenFlagstaff}

Recently fabricated graphene\cite{NovoselovScience04,BergerJPCB04} immediately became a challenge for CR measurements and implied necessity to
comprehend the phenomenon of cyclotron resonance in a system with massless charge carriers.\cite{NovoselovNature05,ZhangNature05}
Special character of CR in graphene can be exemplified on the basis of a simple quasi-classical approach.\cite{ShonJPSJ98,MikhailovPRB09,GusyninNJP09,OrlitaSSC10} If we solve the classical equation of motion for a charged particle with an energy $\varepsilon$ which depends linearly on the momentum $p$, i.e. $\varepsilon = \pm v_{F}|p|$, we find out that the precession-like (cyclotron) motion remains characteristic also for these Dirac-like particles, nevertheless, with a cyclotron frequency modified to $\omega_c=|e|B/(|\varepsilon|/v_F^2)$. Hence, the classical cyclotron motion and consequently also the cyclotron resonance of massless Dirac fermions scale linearly with the applied magnetic field $B$, similar to conventional massive particles, but they depend on the particle energy, which appears in a form of an energy-dependent cyclotron mass defined by the Einstein relation $|\varepsilon|=mv_F^2$. 

Surprisingly, this classical, i.e. linear with the magnetic field, cyclotron resonance has not been clearly observed in graphene so far. Instead, the optical response of graphene has been experimentally explored\cite{SadowskiPRL06,JiangPRL07,DeaconPRB07,OrlitaPRL08II} only in the  quantum regime, when all detected transitions correspond to excitations between well-separated Landau levels (LLs). These measurements provided us with a unique insight into the LL spectrum of graphene and directly visualized its intriguing $\sqrt{B}$-scaling, but no information about the quasi-classical limit could be extracted. Consequently, it remains a challenge to demonstrate experimentally the quasi-classical cyclotron resonance in graphene, or interestingly, to study the crossover  between this classical (linear in $B$) and fully quantum-mechanical (linear in $\sqrt{B}$) magneto-optical response of graphene -- the crossover, which is not present in a system of massive particles, where the response always remains linear in the applied magnetic field. 

Here we report on magneto-optical measurements on highly-doped epitaxial graphene grown on the 
silicon-terminated surface of SiC. We identify the cyclotron-resonance-like response related to the massless 
Dirac fermions in the vicinity of the Fermi level, which evolves nearly linearly with the applied magnetic field,
as expected from (quasi-)classical considerations. These results are compared to the currently well-understood 
response of multilayer epitaxial graphene prepared on the carbon-terminated surface of SiC, where majority 
of (rotationally stacked) graphene sheets is nearly undoped and where the fully quantum-mechanical treatment is appropriate
for its description.

\section{Sample preparation and experimental details}

In our study, we present data taken on two particular samples of graphene prepared by thermal decomposition on the surface 
of semi-insulating silicon carbide.\cite{BergerJPCB04} Both specimen have been grown at 1600$^\circ$C in Ar atmosphere using a commercially available horizontal chemical vapor deposition hot-wall reactor (Epigress V508), inductively heated by a RF generator. 

The sample denoted as A in our study, grown on the C-terminated surface of silicon carbide (4H-SiC[000\={1}]), represents a standard multilayer epitaxial graphene (MEG) specimen. The majority of sheets in this structure is practically undoped (residual $p$-like doping expected at the level of $10^9$~cm$^{-2}$, see Refs.~\onlinecite{OrlitaPRL08II} and \onlinecite{SprinklePRL09}) with the electronic band structure similar if not identical to an isolated graphene monolayer, provided by the characteristic rotational stacking of layers.\cite{HassPRL08} This fact is reflected in the micro-Raman spectra taken on this sample, which revealed dominantly single-component (Lorentzian-like) 2D band,~\cite{FerrariPRL06,FaugerasAPL08} nevertheless, some Bernal-stacked residuals have been observed on selected locations as well.  

The sample B has been prepared on the Si-terminated surface of SiC (4H-SiC[0001]) and also characterized in the
micro-Raman experiment. Similarly to the previous sample, the dominantly observed single-component 2D band serves as an indication 
for linearity of electronic bands. It points towards the presence of a single graphene sheet (in addition to the non-graphene-like buffer layer\cite{MattauschPRL07}), as in contrast to MEG on the C-face, the few-layer graphene stacks on the Si-face exhibit Bernal and not rotational ordering.\cite{OhtaScience06} In addition, such monolayers are typically rather highly doped by the charge transfer from the substrate.\cite{BostwickNaturePhys07,ZhouNatureMater07} 

\begin{figure}
\scalebox{0.55}{\includegraphics*{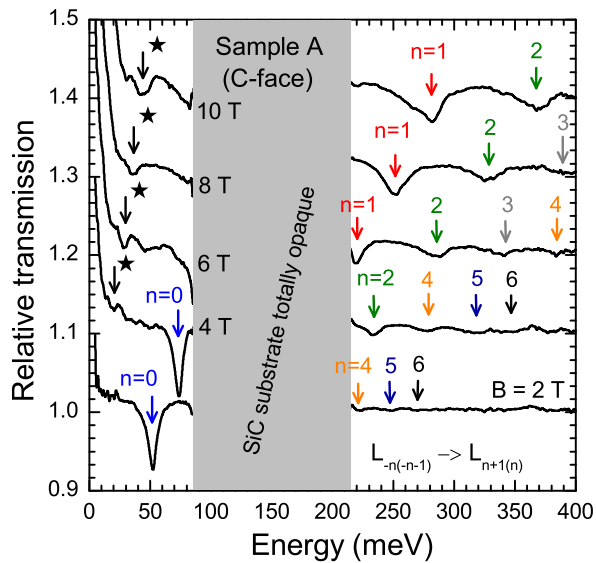}} \caption{\label{SPKT-C-face} (color online) Relative transmission spectra of the sample A (MEG on C-face of SiC) for selected values of the magnetic field. For clarity, successive spectra are shifted vertically by 0.1. Individual transitions discussed in this paper are denoted by vertical arrows.}
\end{figure}

To measure the (non-polarized) infrared transmittance of our samples, we used the
radiation of a globar, which was analyzed by a Fourier transform spectrometer
and delivered to the sample via light-pipe optics. The light was then detected
by a composite bolometer kept at $T=4.2$~K placed directly below the sample.
Measurements were done in the Faraday configuration in a superconducting solenoid. 
Both investigated samples were fully opaque in the energy range 85-220~meV due to phonon-related absorption in SiC.

\begin{figure}
\scalebox{0.75}{\includegraphics*{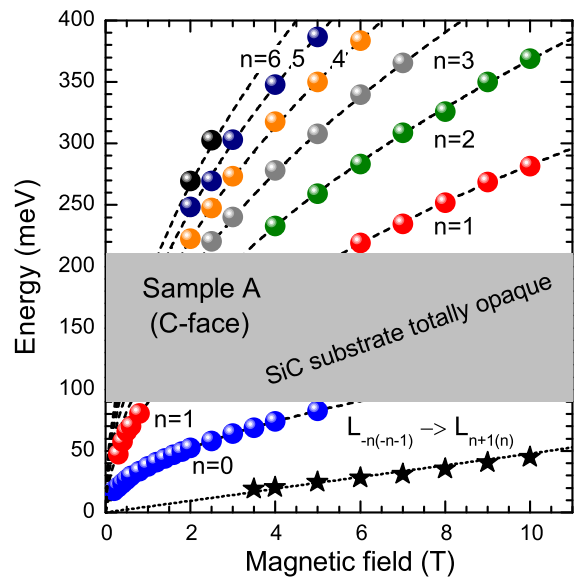}} \caption{\label{FanChart-C-face} (color online) Positions
of absorption lines observed in the A sample grown on C-terminated surface of SiC. The spheres correspond to inter-LL transitions in
electrically isolated and nearly undoped layers in MEG, having a characteristic $\sqrt{B}$-dependence. The dashed lines
show their theoretically expected positions for the Fermi velocity of $v_F=1.02\times10^6$~m.s$^{-1}$. Origin of the transition
denoted by stars, which evolves nearly linearly with $B$, is discussed in the text.}
\end{figure}

\section{Results and Discussion}

Let us start our discussion with the data obtained on the sample A, i.e. on the MEG structure grown on the C-terminated surface of SiC, see Fig.~\ref{SPKT-C-face}, whose magneto-optical response is relatively well understood.\cite{SadowskiPRL06,SadowskiSSC07,OrlitaPRL08II}
Transmission spectra are in this kind of specimens dominated by relatively sharp absorption lines with a typical $\sqrt{B}$-like scaling (see Fig.~\ref{FanChart-C-face}) and taking the LL spectrum of graphene: $E_n=\mathrm{sign}(n)v_F\sqrt{2 \hbar |e Bn|}$, we easily identify them as inter-LL transitions L$_{-n}$$\rightarrow$L$_{n+1}$ and L$_{-(n+1)}$$\rightarrow$L$_n$ with $n=0\ldots6$. The corresponding Fermi velocity can be evaluated as $v_F=(1.02\pm 0.01)\times 10^{6}$~m.s$^{-1}$. As a matter of fact, this response is equivalent to that of an exfoliated graphene monolayer\cite{JiangPRL07,DeaconPRB07,HenriksenPRL10} and in turn it serves as a direct indication that layers in MEG indeed behave as independent, electrically isolated, graphene sheets. 

Apparently, here we deal with graphene in its fully quantum regime, i.e. we study optical excitations between well-separated Landau levels,
whose $\sqrt{B}$-scaling and multi-mode character are directly related to the linearity of the electronic bands.\cite{GusyninPRL07} To characterize our sample further, we can set the upper limit for the carrier density that cannot significantly exceed $10^{10}$~cm$^{-2}$, as the transition L$_{0(-1)}\rightarrow$L$_{1(0)}$ is observed down to the magnetic field of 100~mT (the filling factor $v<6$). We assume that the graphene sheets are slightly $p-$doped as follows from the recent ARPES study done on equivalent specimens.~\cite{SprinklePRL09} We should also note that the dominant graphene-like lines are often accompanied by additional spectral features, which are usually significantly weaker, nevertheless still well-defined in the spectra, located especially at low energies, see e.g. the transition denoted by stars in Fig.~\ref{SPKT-C-face}. We will come back to these features later on after we discuss the response of the sample B. 

The sample B exhibits a very much different behavior in our magneto-transmission experiment and shows no $\sqrt{B}$-scaled spectral features, which are, as we have learned above, characteristic of graphene in its quantum regime. Instead, it clearly shows magnetic-field-induced transmission at low energies accompanied by a relatively wide absorption minimum, which moves towards higher energies with increasing $B$. We found this behavior to be characteristic for all graphene specimens prepared on Si-terminated surface of SiC (not shown in this paper), nevertheless, the total intensity of the observed spectral features as well as the position of the absorption minimum slightly varied from sample to sample. 

\begin{figure}
\scalebox{0.75}{\includegraphics*{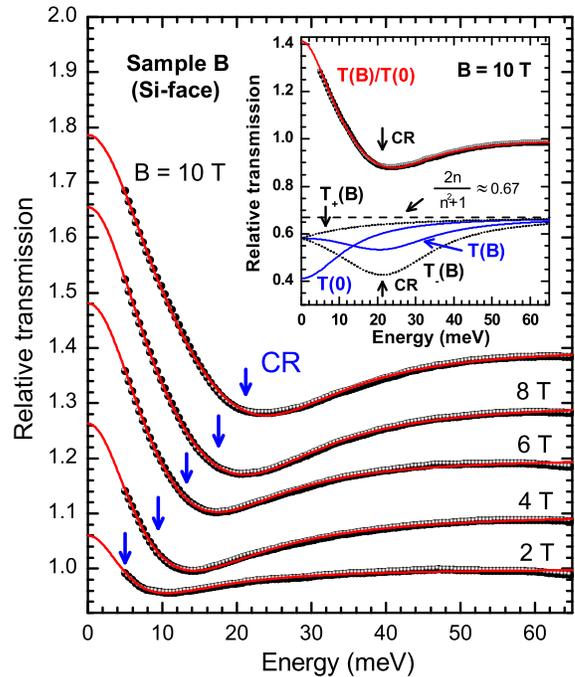}} \caption{\label{SPKT-Si-face} (color online) Transmission spectra of the sample B
taken at various magnetic fields in the far infrared spectral region. Solid curves represent fits to experimental data (spheres) according to the quasi-classical theory described in the text. For clarity, successive spectra are shifted vertically by 0.1. Vertical arrows indicate position of a relatively broad CR resonance, as obtained from the analysis. The inset shows fitting procedure at $B=10$~T in detail. The individual transmission curves $T(0), T(B), T_+(B)$ and $T_-(B)$ have been plotted to demonstrate their contribution to the final relative change of sample transmission $T(B)/T(0)$. Let us note that these curves share a common baseline given by the transmission of the slab without any conducting layer (or away from the resonances) $T_{\mathrm{Slab}}=2n/(n^2+1)\approx0.67$.}
\end{figure}

To explain this behavior, we should recall that epitaxial graphene on the Si-terminated surface is usually highly doped by the charge transfer from the substrate, reaching carrier densities up to $10^{13}$~cm$^{-2}$ ($E_F\approx0.4$~eV), see e.g. Ref.~\onlinecite{BostwickNaturePhys07}, with quality (expressed in terms of scattering time or mobility) comparable to MEG on the C-face of SiC.\cite{EmtsevNatureMater09,OrlitaPRL08II} Therefore, at magnetic fields of a few Tesla, LLs should not be resolved around the Fermi level and the system should remain in the quasi-classical regime. Interestingly, owing to the non-equidistant spacing of LLs in graphene, the tuning from a fully quantum regime to the classical one is possible just by changing the Fermi level.  

If we adopt this, the qualitative explanation of the observed behavior becomes straightforward. The relative magneto-transmission spectrum $T(B)/T(0)$ reflects two specific effects: The low-energy free carrier (Drude-like) absorption, which is gradually suppressed with the increasing field, and the cyclotron resonance 
itself. The former one gives rise to the field-induced transmission at low energies that decreases monotonically with the photon energy, and the latter one is manifested by an absorption minimum close to the energy of the cyclotron resonance $\omega_c$. Let us now proceed with the quantitative analysis of this data, which will allow us to determine the characteristic field-dependence of $\omega_c$. 

In the framework of the classical theory for cyclotron resonance, the optical conductivity is given by 
\begin{equation}
\sigma_{\pm}(\omega,B)=\sigma_0\frac{i\gamma}{\omega\pm\omega_c+i\gamma},
\label{conductivity}
\end{equation}
which describes both the CR active ($-$) and inactive ($+$) polarization modes of light, as well as the Drude-like low-energy
absorption (at $B=0$). $\sigma_0$ here denotes zero-field dc conductivity and $\gamma$ is a phenomenologically introduced damping.  
We should also point out that in graphene, contrary to conventional systems of massive particles, the cyclotron frequency $\omega_c$ is not
independent of $\sigma_0$ and $\gamma$ as the ``effective'' effective mass of Dirac fermions (at the Fermi level) can be determined
from these two parameters.

\begin{figure}
\scalebox{0.42}{\includegraphics*{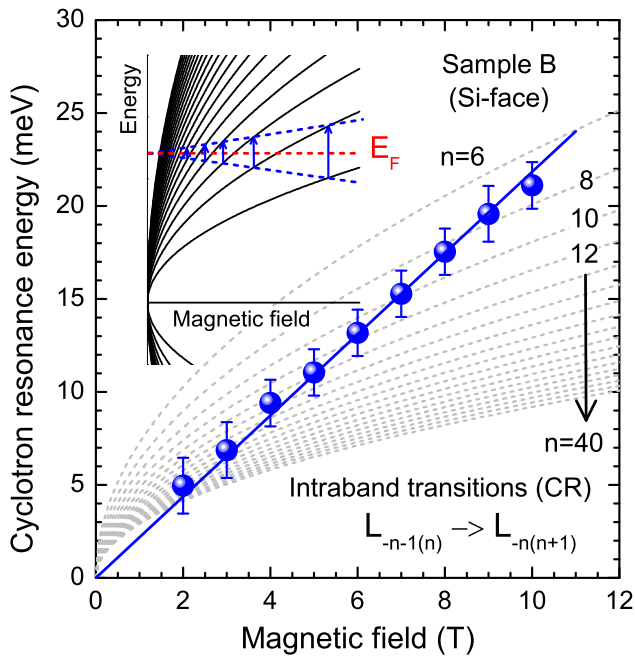}} \caption{\label{FanChart-Si-face} (color online) Position of the cyclotron resonance
in the sample B as extracted from the fitting experimental curves in Fig.~\ref{SPKT-Si-face}, well-following the linear in B dependence plotted for the cyclotron mass of $m=0.053m_0$ (full line). For comparison, we also depict positions of intraband (i.e. cyclotron) resonances between LLs with higher indices, L$_n\rightarrow$L$_{n+1}$ $n=6\ldots40$ (doted lines). The inset schematically shows linear in $B$ evolution of CR in highly doped graphene, 
despite the fact that the individual LLs exactly follow the $\sqrt{B}$-dependence. As we cannot determine the sign of charge of the particles in our non-polarized transmission experiment, we considered $n$-type doping only in this inset for simplicity.}
\end{figure}

The transmission of circularly-polarized light through the studied specimen (graphene sheet on an isolating substrate) can be expressed by formula:\cite{SadowskiIJMPB07}
\begin{equation}
T_{\pm}(\omega,B)=\frac{16n^2}{|\alpha_\pm(\omega,B)|^2-|\beta_\pm(\omega,B)|^2},
\label{Sadowski}
\end{equation}
where
\begin{align*}
\alpha_\pm(\omega,B) & = (n+1)\left(n+1+\frac{\sigma_{\pm}(\omega,B)}{\varepsilon_0 c}\right) \\
\beta_\pm(\omega,B)  & = (n-1)\left(n-1-\frac{\sigma_{\pm}(\omega,B)}{\varepsilon_0 c}\right)	
\end{align*}
which is derived for a dielectric slab with the index of refraction $n$ ($n_{\mathrm{SiC}}\cong2.6$) with an infinitely thin conducting layer on one of its surfaces (described by $\sigma_{\pm}$). $c$ and $\varepsilon_0$ stand for the real speed of light and 
the vacuum permittivity, respectively. The dependence of $T_\pm$ on the thickness of the slab (i.e. interference effects) has been already averaged out.

Within the framework of our simple model, we can easily calculate the relative change of the sample transmission with the magnetic field $[T_+(B)+T_-(B)]/2T(0)$, 
and compare it with our experimental data. As the conductivity as well as the damping parameter can in principle vary with $B$, we have first found 
their appropriate values at zero field, $\sigma_0=(2.4\pm 0.3)\times10^{-3}$~$\Omega^{-1}$ and $\gamma_0=(10 \pm 2)$~meV, which we afterwards kept 
constant and varied only $\omega_c, \sigma_B$ and $\gamma_B$ to reproduce our experimental traces. 

As shown in Figs.~\ref{SPKT-Si-face} and \ref{FanChart-Si-face},
our simple theoretical model is able to reproduce experimental transmission spectra very well and the CR frequency indeed follows a clear linear in $B$ 
dependence. The conductivity $\sigma_B$ and the CR half-width $\gamma_B$ exhibited in our fitting procedure a weak decrease and increase, respectively, but practically
remain within the interval of uncertainty estimated for $\sigma_0$ and $\gamma_0$ and they can be therefore taken as nearly constant in the investigated range of fields. The width (FWHM) 
of the cyclotron resonance absorption peak $2\gamma_B$ has been found comparable with the cyclotron energy $\omega_c$ what justifies the quasi-classical regime in the investigated sample. 

We should also clarify that a relatively broad CR resonance implies that the minimum in the transmission spectra does not exactly match 
to the position of the CR mode. This is explained in the inset of Fig.~\ref{SPKT-Si-face}, where the transmission curves $T(B)$ at $B=10$~T and $T(0)$ are depicted as calculated from parameters obtained in our data analysis. We also plot curves $T_{-}(B)$ and $T_{+}(B)$, which show expected transmission in opposite circular polarizations of light, i.e. in the CR active and inactive modes, respectively. Let us note that all these transmission curves 
share a common baseline at $2n/(n^2+1)\approx 0.67$, which corresponds to the light transmission through a dielectric slab without any conducting sheet (or away from resonances). This value can be easily obtained from Eq.~\eqref{Sadowski} for $\sigma\equiv0$. 

Let us now return to the main finding of our analysis, i.e. to the clear linear in $B$ dependence of the cyclotron resonance as shown in Fig.~\ref{FanChart-Si-face}. If we evaluate the effective mass corresponding to this CR mode, we obtain $m=(0.053\pm0.004)m_0$, which according to the Einstein relation $E_F=mv_F^2$ implies the Fermi level of $E_F\approx310$~meV in a reasonably good agreement with conclusions of ARPES measurements on similar samples.\cite{BostwickNaturePhys07,ZhouNatureMater07} In this estimation, we used the Fermi velocity of $1.02\times10^{6}$~m.s$^{-1}$ determined from the measurement on the sample A, but very similar velocities are reported by ARPES for graphene on the Si-terminated surface of SiC as well.
\cite{BostwickNaturePhys07,ZhouNatureMater07}

The observation of the CR mode that is linear in the applied field $B$ justifies our simple approach in which (otherwise massless) Dirac fermions
are considered as massive particles, the mass of which is determined by the Fermi energy. On the other hand, this confirmation implies another 
constrain on parameters used in our fitting. Namely, the conductivity $\sigma_0$ can be directly estimated within the Boltzmann transport theory from the scattering rate $\gamma_0/\tau$ and from the Fermi energy: $\sigma_0=(2e^2/h)(E_F/\gamma_0)$. This relation was respected in our fitting procedure and required an additional self-consistent loop to keep the correct relation between $\sigma_0$ and $\gamma_0$ on the basis of $E_F$ extracted from the fitting. Obviously, this approach works only in case of specimens with a full coverage by the graphene monolayer, otherwise the conductivity derived from $\gamma_0$ and $E_F$ exceeds the value obtained directly from the fitting. The ratio of these two values thus could serve as a simple parameter that describes fraction of coverage by graphene. 

We should note that the evaluated mass of $0.053m_0$ is surprisingly close the effective mass of $K$ point electrons in bulk graphite $\sim$0.058$m_0$,\cite{Brandt88} but this similarity represents just a coincidence. To prove this, we can for instance argue that bulk graphite reaches the quantum limit at fields around 7~T,\cite{WoollamPRL70,SchneiderPRL09} and therefore it cannot exhibit CR line that corresponds to the effective mass of $K$ point electrons. Instead, due to the formal similarity of the band structure at the $K$ point with the graphene bilayer, it should provide the CR-like absorption at energy by a factor of $\sqrt{2}$  higher, as indeed observed experimentally.\cite{OrlitaPRL09}

To summarize, the highly-doped graphene sheet on the Si-terminated surface allows us to study the CR resonance of masssless Dirac fermions in its fully classical limit, when it becomes strictly linear with the applied magnetic field. Nevertheless, let us look at the same problem using the quantum-mechanical description, i.e. interpret our data as intra-band excitations between adjacent (but in reality significantly broadened) Landau levels, as schematically shown in the inset of Fig.~\ref{FanChart-Si-face}. To clarify the linear in $B$ response, found in a system with a $\sqrt{B}$-scaled LL spectrum, we only need to consider the Fermi level $E_F$ located high in the conduction (or equivalently low in the valence) band and independent of the magnetic field. This can be exactly ensured by the pinning the Fermi level by the surface states in the substrate or fulfilled approximatively,
when LLs significantly overlap. At a given magnetic field, the $n$-th LL approaches the Fermi level, $E_F=E_1\sqrt{n}$, and the active CR-like transition (between adjacent LLs) has the energy of $\hbar\omega_c=E_1(\sqrt{n+1}-\sqrt{n})\approx E_1/(2\sqrt{n})=\hbar e B/(E_F/v_F^2)$, i.e. it indeed becomes linear in $B$ and moves with the slope related to the cyclotron mass $m=E_F/v_F^2$.

The analysis of data obtained on highly-doped graphene in the sample B allows us to come back to the low-energy features in spectra taken on the sample A. Also in this kind of samples, we usually find relatively weak spectral features whose energy moves nearly linearly in $B$. Even though the sample A dominantly consists of nearly undoped graphene sheets, the increase of the magneto-transmission at lower energies serves as an indication of some free carriers in the sample, similarly to the sample B. In fact, also in MEG on the C-face of SiC, first few layers nearby the interface are usually doped by charge transfer from the substrate. The transition denoted by stars (see Figs.~\ref{SPKT-C-face} and \ref{FanChart-C-face}) thus could be assigned to the CR response of these doped layers, and its characteristic slope $\approx$4.5~meV.T$^{-1}$ (effective mass of $0.025m_0$) would imply the Fermi level of 
$|E_F|\approx150$~meV. On the other hand, the carrier density in these doped layers gradually decreases with the distance from the substrate, which should 
prevent observation of a well-defined CR mode (whose energy depends on the density of massless particles), while the transition denoted by the star remains relatively narrow in our data. 

More likely, this line originates in Bernal-stacked residuals, especially in the graphene bilayer which has been identified as a minor but 
not negligible component of MEG.\cite{SiegelPRB10,Orlita10}
The masssive Dirac fermions in the bilayer are characterized by an effective mass of $m_B=\gamma_1/(2v^2_F)\approx0.32m_0$  ($\gamma_1=380$~meV\cite{HenriksenPRL08,ZQLiPRL09,KuzmenkoPRB09,Orlita10}) which does not match to the effective mass derived for the ``star'' transition, but
the characteristic LL spectrum in the bilayer, $E_n\cong \mathrm{sign}(n)(\hbar |e B|/m_B)\sqrt{|n|(|n|+1)}$,\cite{McCannPRL06,AbergelPRB07} implies in the 
quantum limit appearance of the CR mode (L$_{-1(0)}\rightarrow$L$_{0(1)}$ transitions) whose slope is by a factor of $\sqrt{2}$ higher. The transition denoted by the star, with a slope corresponding to the effective mass of $0.025m_0$, thus could be assigned an undoped graphene bilayer in which the CR absorption
should correspond to the mass of $m_B/\sqrt{2}\approx0.023m_0$. Nevertheless, the definitive conclusion cannot be drawn on the basic of our current data. 

\section{Conclusions}

We have explored magneto-optical response of graphene in two distinctively different regimes. 
Whereas the weakly doped sheets in multilayer epitaxial graphene on the C-terminated surface allowed us to study 
the optical response of graphene in the quantum limit, massless Dirac fermions in the quasi-classical regime 
can be explored in highly doped epitaxial graphene prepared on the Si-terminated surface of SiC. 
The response of the former system clearly shows $\sqrt{B}$-like scaling, typical of Landau 
levels in systems with linear electronic bands, the latter case proves that cyclotron resonance of massless particles 
remains in the quasi-classical limit linear with the magnetic field. 

\vspace{0.2cm}
\noindent\emph{Additional note:} During our work on the manuscript we became aware of Faraday rotation measurements on epitaxial graphene,\cite{CrasseeCM10} 
in which a similar conclusion about cyclotron resonance of massless Dirac fermions in the quasi-classical limit has been made. 

\begin{acknowledgments}
This work has been supported by the projects 395/N-PICS-FR/2009/0 and projects MTKD-CT-2005-029671, furthermore by
grants 670/N-ESF-EPI/2010/0, 671/N-ESF-EPI/2010/0 and GRA/10/E006 within the ESF EuroGraphene programme (EPIGRAT). 
We acknowledge also funding received from EuroMagNETII under the EU contract No.~228043 and via projects MSM0021620834 and GACR No.~P204/10/1020.
\end{acknowledgments}

%\bibliography{SiC-Andrzej}

\end{document}